\documentclass[twocolumn,showpacs,preprintnumbers,amsmath,amssymb,pra,unsortedaddress,superscriptaddress,longbibliography,letterpaper]{revtex4-1}
\usepackage{graphicx}
\usepackage{algorithmicx}
\usepackage{algorithm}
\usepackage{algpseudocode}
\usepackage{xfrac}

\begin{document}


\title{Perturbed Field Ionization for Improved State Selectivity}

\author{Vincent C. Gregoric}%
\affiliation{Department of Physics, Bryn Mawr College, Bryn Mawr, PA 19010.}
\affiliation{Department of Physics, Austin College, Sherman, TX 75090.}

\author{Jason J. Bennett}
\affiliation{Department of Physics and Astronomy, Ursinus College, Collegeville, PA 19426.}

\author{Bianca R. Gualtieri}
\affiliation{Department of Physics and Astronomy, Ursinus College, Collegeville, PA 19426.}
\affiliation{Department of Physics, Temple University, Philadelphia, PA 19122.}

\author{Hannah P. Hastings}%
\affiliation{Department of Physics, Bryn Mawr College, Bryn Mawr, PA 19010.}

\author{Ankitha Kannad}%
\affiliation{Department of Physics, Bryn Mawr College, Bryn Mawr, PA 19010.}

\author{Zhimin Cheryl Liu}%
\affiliation{Department of Physics, Bryn Mawr College, Bryn Mawr, PA 19010.}
\affiliation{Department of Physics, University of Colorado, Boulder, CO 80309.}

\author{Maia R. Rabinowitz}%
\affiliation{Department of Physics, Bryn Mawr College, Bryn Mawr, PA 19010.}

\author{Zoe A. Rowley}
\affiliation{Department of Physics and Astronomy, Ursinus College, Collegeville, PA 19426.}

\author{Miao Wang}%
\affiliation{Department of Physics, Bryn Mawr College, Bryn Mawr, PA 19010.}

\author{Lauren Yoast}
\affiliation{Department of Physics and Astronomy, Ursinus College, Collegeville, PA 19426.}

\author{Thomas J. Carroll}
\affiliation{Department of Physics and Astronomy, Ursinus College, Collegeville, PA 19426.}

\author{Michael W. Noel}%
\affiliation{Department of Physics, Bryn Mawr College, Bryn Mawr, PA 19010.}

\date{\today}

\begin{abstract}
Selective field ionization is used to determine the state or distribution of states to which a Rydberg atom is excited.  By evolving a small perturbation to the ramped electric field using a genetic algorithm, the shape of the time-resolved ionization signal can be controlled.  This allows for the separation of signals from pairs of states that would be indistinguishable with unperturbed selective field ionization.  Measurements and calculations are presented that demonstrate this technique and shed light on how the perturbation directs the pathway of the electron to ionization. Pseudocode for the genetic algorithm is provided. Using the improved resolution afforded by this technique, quantitative measurements of the $36p_{3/2}+36p_{3/2}\rightarrow 36s_{1/2}+37s_{1/2}$ dipole-dipole interaction are made.
\end{abstract}


\maketitle

\section{Introduction}

Selective field ionization (SFI) allows for measurements of the population distribution in atoms excited to Rydberg states.  In this technique, a ramped electric field is applied to the Rydberg sample, which ionizes the atoms, and the resulting time-resolved signal reveals the state distribution \cite{gallagher_rydberg_1994}.  Early features in the signal are associated with weakly bound electrons (high $n$) that ionize at low field while late features result from tightly bound electrons (low $n$) that ionize at high field.  Such measurements have been used in a broad array of research including studies of quantum-classical correspondence with atomic electron wave packets \cite{noel_excitation_1996,jones_ramsey_1993,preclikova_excitation_2012}, strong-field multiphoton interactions~\cite{story_resonant_1993,gatzke_short-pulse_1994,hoogenraad_far-infrared_1995,gatzke_quantum_1995,conover_chirped-pulse_2002}, quantum control using optical \cite{kozak_state-selective_2013,noel_shaping_1997}, terahertz \cite{jones_measurement_1999,mandal_half-cycle-pulse-train_2010}, and microwave fields \cite{bayfield_excited_1988,noel_population_1999}, ultracold plasmas \cite{killian_formation_2001,crockett_heating_2018}, long-range dipolar interactions \cite{van_ditzhuijzen_spatially_2008,altiere_dipole-dipole_2011,richards_dipole-dipole_2016,yakshina_line_2016,lee_excitonlike_2017}, and many more.  

While the simple picture of SFI described above is generally correct, the details of the field ionization process can significantly complicate the time-resolved ionization signal.  The Stark effect splits and shifts states of different angular momentum.  As the electric field increases, a particular state will encounter and interact with hundreds of other states before being ionized.  Early experiments classified field ionization as either adiabatic or diabatic.  In sodium and lithium it was observed that population jumping across avoided crossings along a diabatic pathway ionized much later in time than population which took the adiabatic pathway~\cite{jeys_diabatic_1980,noel_classical_2000}.  In fact, the wide range of avoided crossing sizes that is encountered also leads to significant broadening of the ionization signal even if the first few avoided crossings are traversed in a predominantly adiabatic or diabatic fashion.  Ultimately, this can make it difficult to distinguish the signals from closely spaced energy eigenstates.

Recently a modification to SFI was developed that allows for coherent manipulation of the time-resolved ionization signal \cite{gregoric_quantum_2017}.  In this technique, a small perturbation to the field ionization ramp is evolved using a genetic algorithm (GA) to produce the desired time-resolved ionization signal.  The small perturbation allows one to direct the pathway of the electron through the Stark map on the way to ionization so this technique is referred to as directed field ionization (DFI).  The ability of the GA to optimize perturbations to achieve several basic fitness goals is presented in \cite{gregoric_quantum_2017}.

DFI has also been used to improve the selectivity of experiments in which the electron is initially in a superposition of two states.  In this case, the same perturbation must direct each state to ionize at a different time.  Three different fitness scores were evaluated for their ability to achieve this signal separation in \cite{gregoric_improving_2018}.

In this paper, we briefly review DFI and present a series of calculations and experiments that give us further insight into this quantum control technique.  We then present an example of how this technique can be used to separate the signals from two states that are significantly overlapped in the SFI signal with a new fitness score that allows for precise quantification of the state distribution.  A short electric field glitch is used to probe the Stark map for these states revealing the most important regions available for optimization.  Finally, we make use of DFI to probe the $36p_{3/2}+36p_{3/2}\rightarrow 36s_{1/2}+37s_{1/2}$ dipole-dipole interaction among Rydberg atoms.  Pseudocode for our GA is included in the appendix.  

\section{Quantum Control of Selective Field Ionization}

These experiments were done in a magneto-optical trap (MOT) where roughly $10^6$ rubidium-85 atoms are cooled to a temperature of 200~$\mu$K.  Rydberg states are excited using additional diode lasers.  The Rydberg atom sample is located between a set of cylindrical electrodes with which both static and time varying electric fields can be applied.  

\begin{figure}
	\centering
	\includegraphics{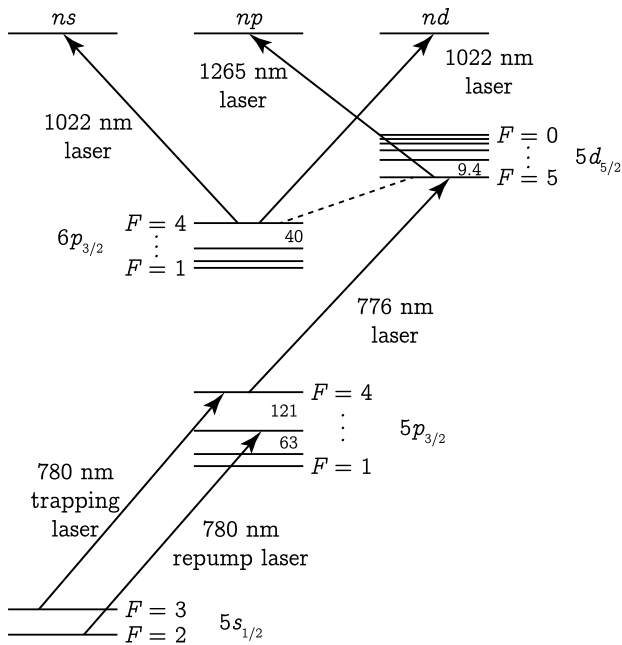}
	\caption{
	Rydberg state excitation. Two lasers operating at 780~nm are used for laser cooling and trapping of the rubidium-85 atoms.  From the $5p_{3/2},\ \ F=4$ state, a 776~nm laser excites the $5d_{5/2}$ state.  This laser is tuned so that atoms are predominantly excited to the $F=5$ state, but some $F=4$ atoms are also excited.  From the $5d_{5/2}$ state, atoms can spontaneously decay to the $6p_{3/2}$ state.  Most atoms end up the $F=4$, but some fall to the $F=3$ state.  $ns$ or $nd$ states are excited with a 1022~nm laser and $np$ states with a 1256~nm laser.
	}
	\label{fig:excitationoutline}
\end{figure}

Our Rydberg state excitation scheme is shown in Fig.~\ref{fig:excitationoutline}~\cite{fahey_excitation_2011}.    With this system we can excite $ns$, $np$, or $nd$ states and our line width is narrow enough to resolve the fine structure splittings of these states.  With the addition of a small static electric field, we can split and therefore select the desired $|m_j|$ state.  The 776~nm laser is pulsed (10~$\mu$s width at 60~Hz) using an acousto-optic modulator so that the momentum transfer from this beam does not remove atoms from the trap.  The 1022~nm and 1265~nm lasers are also pulsed using acousto-optic modulators in a double pass configuration.  This allows us to tune the lasers on or off resonance and alternate between exciting two states of different $|m_j|$ or $\ell$.  The 776~nm and Rydberg lasers are focused to a spot size of $\sim100\ \mu$m.  By sending the 776~nm and Rydberg beams into the vacuum system from orthogonal directions, we are able to excite a small volume of trapped atoms, which reduces the effect of any field inhomogeneities.  

\begin{figure}
	\centering
	\includegraphics{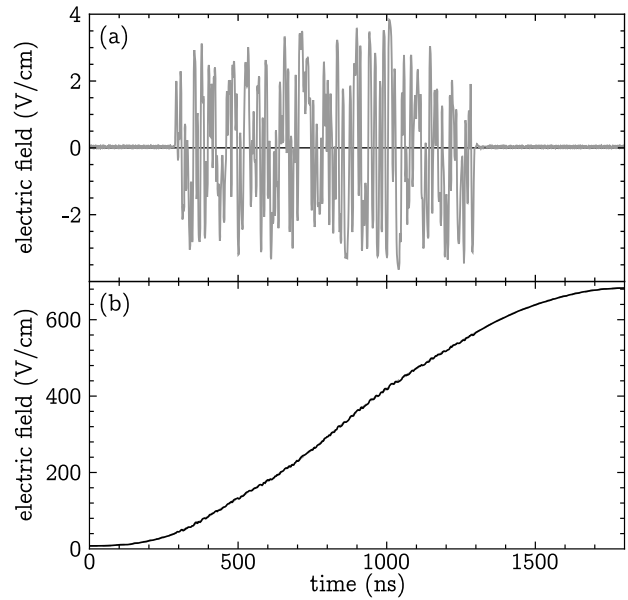}
	\caption{
	Applied electric fields.  (a) Perturbing field. The size of the perturbations is quite small compared to the high voltage ionizing ramp but is sufficient to sweep through Stark avoided crossings multiple times. A typical perturbation consists of about 1000 voltage values. (b) Combined ionizing and perturbing electric field seen by the Rydberg atoms. Variations to the ramp are barely visible on this scale.
	}
	\label{fig:FIP_arb_total}
\end{figure}

Initially we focus on controlling the ionization signal from a single state.  To do this we turn off the 1265~nm laser and tune the 1022~nm laser to excite atoms to the $32d_{5/2,|m_j|=3/2}$ state, which is separated from the other $|m_j|$ states by applying a small static electric field ($\approx 3$~V/cm).  To the electrode on one side of our MOT we apply a voltage that ramps to a peak value of 1800~V in 1.5~$\mu$s.  To the electrode on the opposite side we apply a small perturbing voltage whose shape can be varied with an arbitrary waveform generator.  A typical perturbation consists of about 1000 voltage values at 1~ns resolution. The resulting electric field seen by the atoms is shown in Fig.~\ref{fig:FIP_arb_total}.  The perturbing field allows us to direct the pathway of the electron through the complicated set of avoided crossings that are encountered on the way to ionization.  Given the complexity of the Stark map as well as our inability to completely characterize the experimental conditions, it is impossible to calculate the perturbing field that will yield the desired result.  We therefore employ a GA to find a perturbation that best achieves the desired field ionization signal.  

An overview of the algorithm is shown in Fig.~\ref{fig:GAloopoutline} and pseudocode can be found in the appendix.  First, an initial population of randomly chosen perturbations is created. A time-resolved ionization signal is collected for each perturbation. A fitness score is calculated for each perturbation based on how well it achieves the target signal shape. The fittest members of the population are mated to form a new population.  Some perturbations are randomly mutated and the cycle is then repeated with the new population.

\begin{figure}
	\centering
	\includegraphics{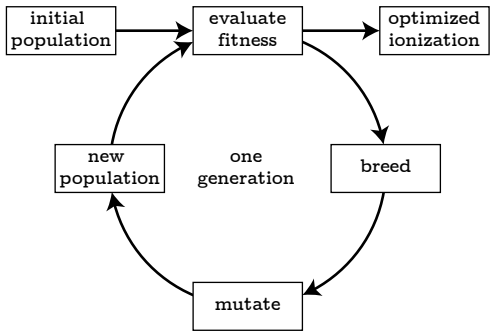}
	\caption{
	Simplified flow chart of the GA. A randomly generated initial population of electric field perturbations is scored based on how well their associated field ionization signals match our goal. The best performing population members breed to create new child perturbations which may be mutated before being placed into a new generation. Data is collected with the new population and the cycle repeats for a fixed number of generations or until some target optimization is achieved. 
	}
	\label{fig:GAloopoutline}
\end{figure}

\begin{figure}
	\centering
	\includegraphics{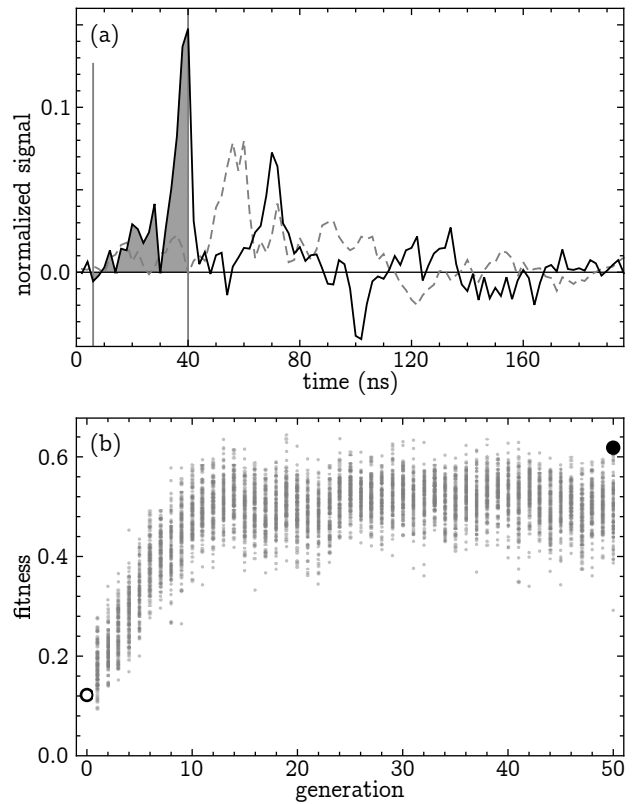}
	\caption{
	Example of experimental GA optimization. The $32d_{5/2,|m_j|=3/2}$ state is initially excited and the GA attempts to move as much of the field ionized signal as possible into the gated region at early time. (a)~The field ionization signal of the best performing optimization (solid) and the unperturbed ionizing ramp (dashed). The fitness score is simply the fraction of the signal that arrives in the gate illustrated by the vertical lines at 5~ns and 40~ns. This optimization was able to put the shaded 60\% of the signal into the gated region. (b)~Fitness scores as a function of generation. The score for the unperturbed ramp is shown by the open circle and the best optimization is shown by the solid black circle at the 50$^{th}$ generation. The fitness score increases rapidly and then plateaus after about 15 generations.
	}
	\label{fig:Example_Optimization}
\end{figure}

An example optimization is shown in Fig.~\ref{fig:Example_Optimization} for field ionization of the $32d_{5/2,|m_j|=3/2}$ state.  Here the goal was to move the time-resolved ionization signal to the region between the two vertical lines in Fig.~\ref{fig:Example_Optimization}(a).  The dashed line shows the time-resolved signal with no perturbation applied, that is, the standard SFI signal.  The solid curve shows the field ionization signal achieved with the best performing perturbation after 50 generations of evolution.  Roughly 60\% of the signal now lies in the desired region, which is shaded in gray.  Fitness scores calculated for each generation are shown in Fig.~\ref{fig:Example_Optimization}(b).  The open circle shows that without the perturbation about 12\% of the signal is in the desired region.  The fitness score increases rapidly at first and then plateaus for the remainder of the optimization.

We have also simulated the example optimization of Fig.~\ref{fig:Example_Optimization} using previously developed calculations~\cite{feynman_quantum_2015,gregoric_quantum_2017,gregoric_improving_2018}. In short, we use a parallel supercomputer to simultaneously time evolve each member of our population. Solving the time-dependent Schr\"odinger equation allows us to track the amplitude and phase of each Stark state as the field increases to ionization.  The resulting calculated field ionization signals are evaluated using the same fitness score and evolved with the same GA as in the experiment. The calculations presented here use a basis of size 629 surrounding the $32d_{5/2,|m_j|=3/2}$ state. Continuum states are not included in the basis; rather, ionization rates are calculated in the parabolic basis using a semi-empirical formula~\cite{damburg_hydrogen_1979}. The time step of our calculation is 0.01~ns, which is small enough for our calculated signals to have converged.

The results are remarkably similar to Fig.~\ref{fig:Example_Optimization}. Our best calculated optimization yields about 65\% of the signal in the gate region and the GA plateaus after 15-20 generations. Of course, the perturbations produced by the simulated optimization are not experimentally useful. One of the advantages of the GA is that it automatically takes into account incompletely characterized factors such as stray fields or inhomogeneous electric and magnetic fields. However, the simulation does allow us to track the electron's path to ionization, information that is not experimentally accessible.

\begin{figure*}
	\centering
	\includegraphics{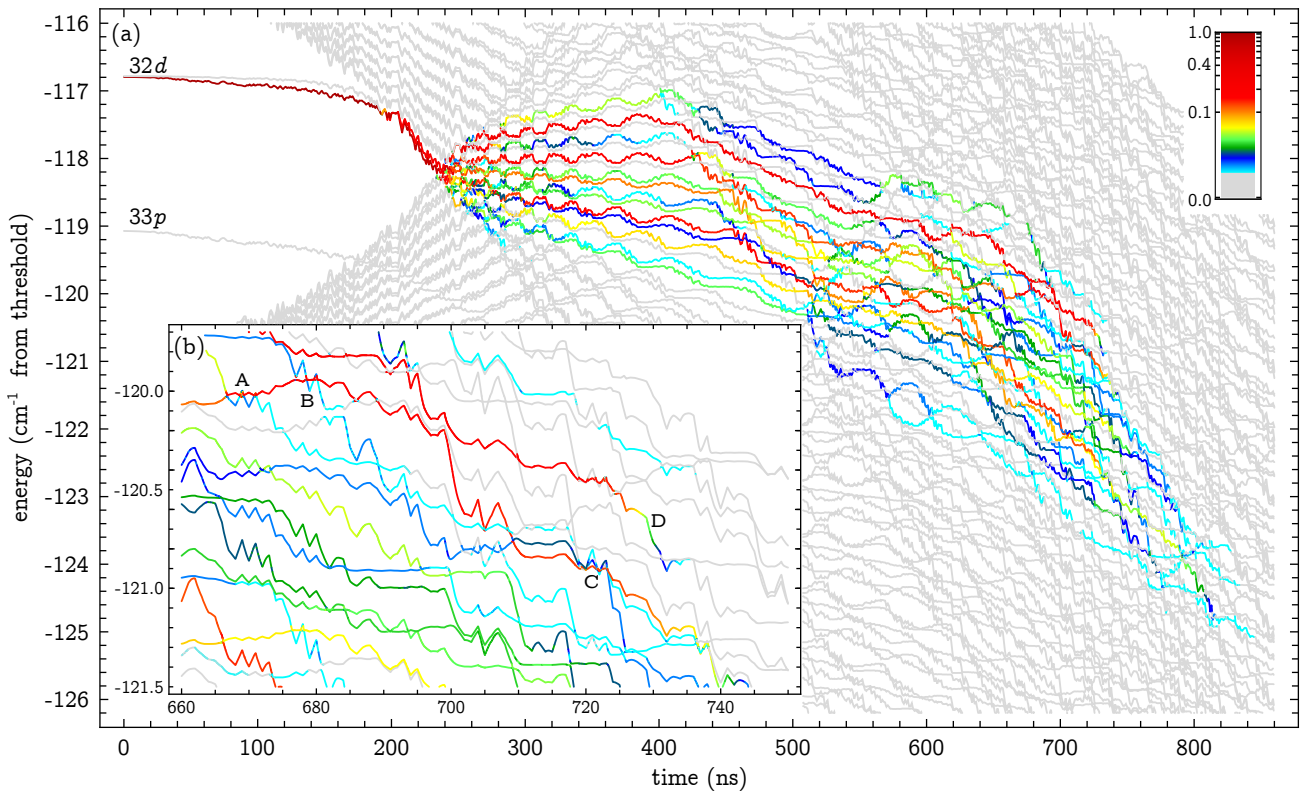}
	\caption{
		(Color online) Simulated GA optimization for the same fitness score used in Fig.~\ref{fig:Example_Optimization}, which is attempting to move the ionization signal to the gated region at early time. The Stark map is graphed as a function of time by using the optimized field ionization ramp to convert from electric field to time. The color of each line is determined by the electron amplitude. The perturbations can sweep back and forth in field as the ramp increases as shown in Fig.~\ref{fig:FIP_arb_total}. Thus, individual avoided crossings in the Stark map can become sets of multiple avoided crossings here. (a) The electron's optimized path to ionization. Near where the manifolds collide at around 250~ns, the GA acts to push population to higher energy and generally early ionizing states. (b) A region near ionization highlighting the ionization of two relatively highly populated Stark lines at \textbf{C} and \textbf{D}. At the avoided crossings \textbf{A}, \textbf{B}, and \textbf{C} the GA guides the population to stay in one particular Stark line because after \textbf{C} that line ionizes rapidly compared to other nearby lines. Similarly, the other highly populated line ionizes even more rapidly at \textbf{D}.
	}
	\label{fig:SimulatedGA}
\end{figure*}

The electron's path to ionization for the best simulated optimization is shown in Fig.~\ref{fig:SimulatedGA}. The Stark map in Fig.~\ref{fig:SimulatedGA} is graphed using the optimized field ionization ramp to convert from electric field to time along the horizontal axis. This shows directly that individual avoided crossings are traversed multiple times as the perturbing voltage sweeps back and forth, as in Fig.~\ref{fig:FIP_arb_total}(a). For example, the set of avoided crossings labeled \textbf{A}, \textbf{B}, and \textbf{C} in Fig.~\ref{fig:SimulatedGA}(b) each represent an individual avoided crossing that is being traversed multiple times by the optimized ramp. Multiple traversals allow coherent interference effects to transfer more population to a particular state than might otherwise be possible with available slew rate adjustments.

A useful analogy is a light wave interacting with a series of beam splitters.  The reflectivity of a particular beam splitter is controlled by the slew rate through the avoided crossing.  In the case where the optimization would benefit from transferring more population to a particular state than is possible with a simple slew rate adjustment, it can traverse the same avoided crossing many times to improve the transfer. For example, if the maximum split at an avoided crossing is 90\% of the population in one state and 10\% in the other, then a second pass of this crossing at the same rate could put all of the population back in the original state or transfer 36\% of it to the other state. Here the phase evolution between traversals of this avoided crossing is critical in determining the final transfer.

\section{Physical Insight into the Ionization Optimization \label{sec:insight}}

Our GA optimizes the electric field perturbations using only the final ionization signals; it is ignorant to the details of ionization. However, the optimized perturbations may yield some physical insight into the ionization process. The results of a single optimization are not particularly illuminating, since the best-performing perturbation typically appears to be a random series of voltages (see Fig.~\ref{fig:FIP_arb_total}(a)). In order to learn more, we used the GA to evolve several solutions under the same conditions, allowing us to compare multiple optimized perturbations.

\begin{figure}
	\centering
	\includegraphics{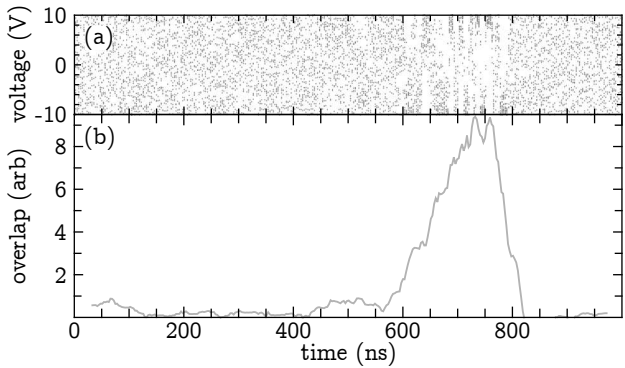}
	\caption{(a)~Scatter plot of the best perturbations from 20 identical GA runs. (b)~Moving overlap integral of the perturbations from (a). This was calculated by taking the average overlap integral of all pairwise combinations of the perturbations over sequential time windows of width 60~ns.}
	\label{F_DFI_repeated_scatter}
\end{figure}

In the first of these experiments, we ran the GA 20 times using the $32d_{5/2,\left|m_j\right|=3/2}$ state. In these scans, all experimental parameters (e.g., mutation rate, target gate, etc.) were held constant. Due to the random nature of choosing an initial population, breeding, and mutation, the best-performing perturbations from these 20 runs are not identical. However, patterns emerge when comparing the optimized perturbations to one another. This can be seen in Fig.~\ref{F_DFI_repeated_scatter}(a), which is a scatter plot showing all 20 optimized perturbations.  Each dot in this plot is the optimized voltage value for that particular time for each of the 20 optimizations. This plot appears quite random, except for the region from 600 to 800~ns. The existence of a regular pattern here indicates consistency between different optimized perturbations. 

This regularity is also apparent in Fig.~\ref{F_DFI_repeated_scatter}(b), which is a ``moving overlap integral'' of the best-performing perturbations. To generate this plot, the overlap integral was calculated for all possible pairwise combinations of perturbations over the first 60~ns, and the results were averaged. This was then repeated as the 60~ns time window was shifted throughout the whole time range. The resulting average overlap integral values are plotted in Fig.~\ref{F_DFI_repeated_scatter}(b) as a function of the center of the overlap integral's time window. The large peak between 600~ns and 800~ns indicates that the best-performing perturbations are most similar in this time range. For this experiment, the region in time between 600~ns and 800~ns corresponds to the portion of the Stark map just before and during ionization (the far right of Fig.~\ref{fig:SimulatedGA}(a)). This suggests that the GA is converging to a preferred solution in this region.

\begin{figure}
	\centering
	\includegraphics{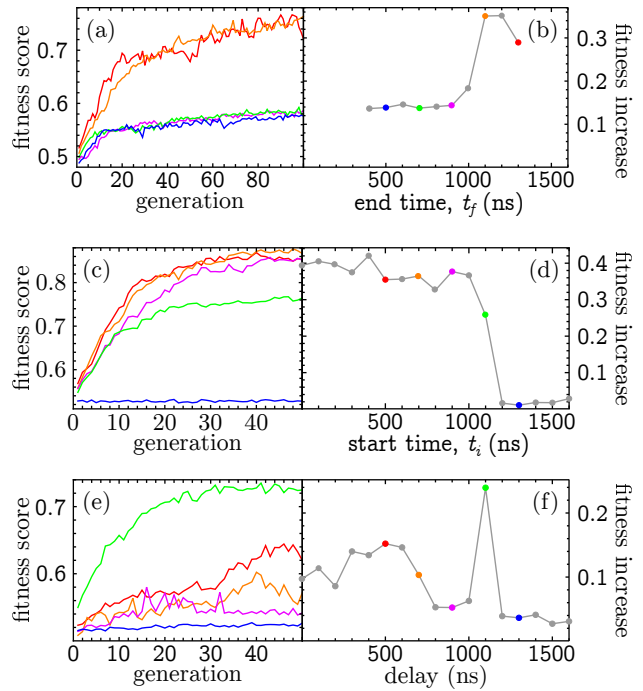}
	\caption{(Color online) Repeated GAs with changing perturbation timing. The plots in (a),~(c), and~(e) show the average fitness score as a function of generation for various perturbation timings, using the same colors as the corresponding graph on the right. The plots in (b),~(d), and~(f) show the increase in fitness score between the unperturbed case and the maximum fitness score in the final generation for various perturbation timings. In (a)~and~(b) the perturbation end time $t_f$ is changed between GA runs while the perturbation start time $t_i$ remains fixed. This is reversed in (c)~and~(d). For the plots in (e)~and~(f), the \textit{length} of the perturbation is held constant while $t_i$ and $t_f$ are changed together, varying the delay between the start of the field ionization ramp and the perturbation.}
	\label{F_DFI_changing_arb_length}
\end{figure}

While the avoided crossings near ionization seem to be important, other avoided crossings may also play a significant role in changing the shape of the signal. To explore this possibility, we have repeatedly run the GA with parameters which are identical except for the timing of the perturbation. This allows us to isolate the influence of different groups of avoided crossings. All experimental parameters were constant except for the start and end times of the perturbation ($t_i$ and $t_f$, respectively). The results of these experiments are shown in Fig.~\ref{F_DFI_changing_arb_length}. In all of these plots, the start of the SFI ramp corresponds to $t$~=~0, while ionization occurs somewhere between 1000~ns and 1100~ns. Note that in this study we have decreased the slew rate of our ionizing ramp which shifts ionization later in time compared to Fig.~\ref{F_DFI_repeated_scatter}. 

For the data in Fig.~\ref{F_DFI_changing_arb_length}(a) and~(b), $t_f$ is different in each GA run, while $t_i$ is held constant. Fig.~\ref{F_DFI_changing_arb_length}(a) shows the average fitness score as a function of generation for five different values of $t_f$. The \textit{increase} in fitness score between the unperturbed case and the best-performing perturbation is shown as a function of $t_f$ in Fig.~\ref{F_DFI_changing_arb_length}(b), using the same colors as in Fig.~\ref{F_DFI_changing_arb_length}(a). When the end of the perturbation occurs after ionization, the fitness increase is $\approx$0.35. But if $t_f$ is decreased to 1000~ns or less, the fitness increase drops off to $\approx$0.15. Note that the fitness increase does not go down to zero, signifying that the early avoided crossings also contribute to the GA's ability to change the signal shape.

Fig.~\ref{F_DFI_changing_arb_length}(c) and~(d) show a similar study, this time with $t_f$ fixed and $t_i$ varied. As long as the perturbation begins before ionization, there is a significant fitness increase ($\approx$0.4). Once \mbox{$t_i > 1100$ ns}, the fitness increase drops off to zero, since the perturbation doesn't begin until after the Rydberg electrons have been ionized and detected.

In Fig.~\ref{F_DFI_changing_arb_length}(e) and~(f), the \textit{length} of the perturbation is held at a constant 100~ns, but the \textit{delay} with respect to the start of the SFI ramp is changed (in other words, $t_i$ and $t_f$ are both changed by the same amount). There is a fitness increase ($\approx$0.12) for perturbations in the range 0 to 700~ns. This decreases for perturbations from 800 to 1000~ns. The large spike at 1100~ns corresponds to when the perturbation occurs during ionization. After this, the fitness increase drops off again, since the electrons are already ionized before the perturbation starts. 

This is consistent with our earlier conclusions from Fig.~\ref{F_DFI_repeated_scatter} and Fig.~\ref{F_DFI_changing_arb_length}(a)~--~(d). The early perturbations are able to control the pathway through the avoided crossings in the region after the manifolds collide and are able to significantly ``deflect'' the electron's subsequent pathway. This can produce a significant increase in fitness score. The perturbation that occurs during ionization produces an even larger increase in fitness by selecting between more or less rapidly ionizing states. The perturbations that occur at intermediate times are not able to use either strategy and produce only minimal fitness score increases.

Physically, the GA has more success when using the avoided crossings in the vicinity of ionization because of the widely varying ionization rates of neighboring states, as discussed previously in \cite{gregoric_improving_2018}. States where the Rydberg electron is concentrated on the up-field side of the ionic core (``red'' states) are relatively easily ionized, while states with the Rydberg electron concentrated on the down-field side of the core (``blue'' states) are harder to ionize. The ionization rates of such states may differ by orders of magnitude.

Due to the Stark effect, red and blue states are coupled, exhibiting avoided crossings. The GA can use crossings between red and blue states near ionization to great advantage when trying to manipulate the shape of the ionization signal. If it is desirable for electrons to be ionized at a certain time in the signal, the GA will favor perturbations which transfer population from blue states to red states at that time. Conversely, switching population from red states to blue states allows the GA to postpone ionization to a later time. 

We can see this behavior directly in our simulation. A small region of the electron's optimized path to ionization is shown in Fig.~\ref{fig:SimulatedGA}(b), highlighting the ionization of two relatively highly populated Stark lines at \textbf{C} and \textbf{D}. The lower in energy of these two lines is populated at \textbf{A}, where two lower population lines meet in a series of avoided crossings. One can see by scanning down from \textbf{A} that these avoided crossings are not significant for other sets of Stark lines and that, in fact, most of the avoided crossings similar in size to those at \textbf{A} are traversed diabatically. At \textbf{A}, the GA has adjusted the phases and slew rates so as to combine the population of two lines almost entirely into just one line.

Following this higher population line through increasing time, we see similar avoided crossings at \textbf{B} and \textbf{C} which preserve the bulk of the population in this one particular Stark line. After \textbf{C} it is clear why the GA has evolved this perturbation to do so. Since the goal of this optimization is to maximize the signal in an early gate, it is advantageous to put population in rapidly ionizing states. In the approximately 20~ns after \textbf{C}, most of the population in this Stark line ionizes (an ionization rate of $10^7$-$10^8$~s$^{-1}$). However, the population in the two Stark lines directly below barely changes during this time (an ionization rate closer to $10^5$~s$^{-1}$). The other highly populated line in Fig.~\ref{fig:SimulatedGA}(b) ionizes even more rapidly at \textbf{D}. Both of these Stark lines ionize within the target gate.

\section{Separating Ionization Signals from Nearby States \label{sec:GAsp}}

\begin{figure}
\centering
\includegraphics{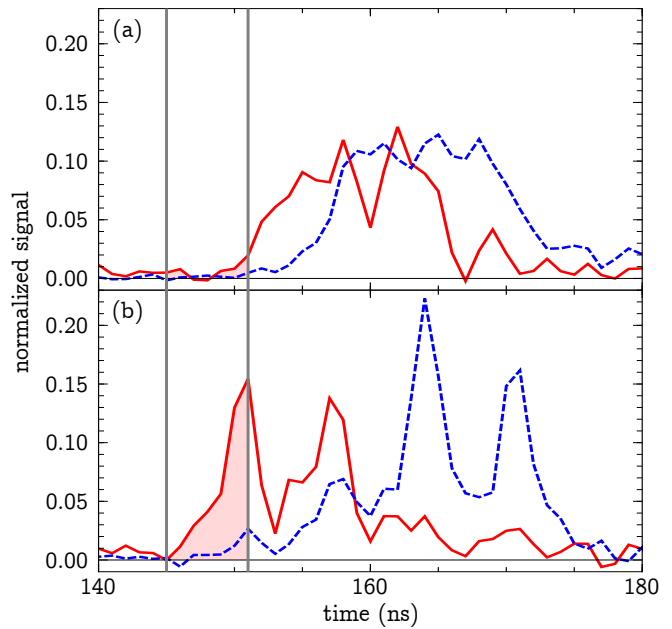}
	\caption{(Color online) Time-resolved ionization signals of the $37s_{1/2}$ (solid, red) and $36p_{3/2}$ (dashed, blue) states for the (a) unperturbed field ionization ramp and (b) the optimized ramp. The goal of the GA was to maximize the $s$ signal while minimizing the $p$ signal in the region indicated by the vertical gray lines. After the GA has optimized the electric field perturbation, the fraction of the $s$ signal within the gate increases to 27.2\% while only 2.1\% of the $p$ signal ends up in this region.}
	\label{fig:GAsp}
\end{figure}

\begin{figure*}
	\centering
	\includegraphics{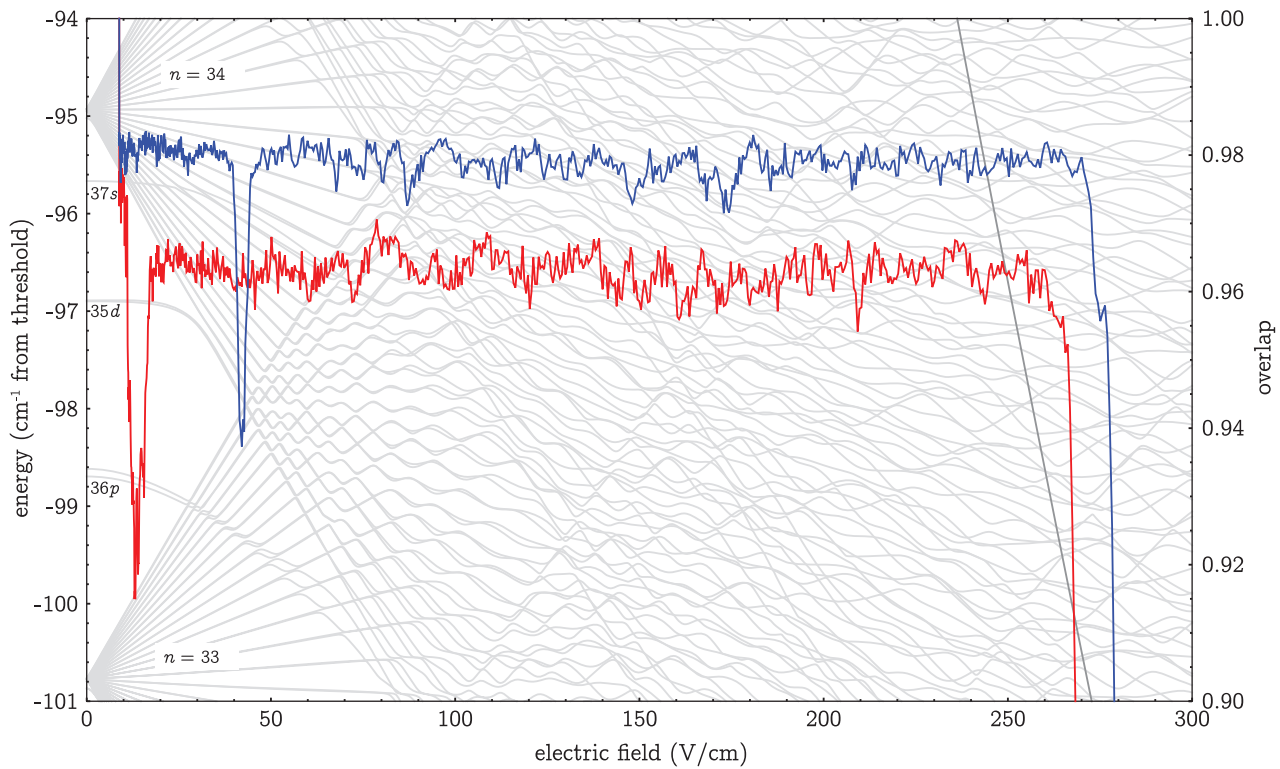}
	\caption{(Color online) Glitch scan.  The time delay of an electric field step is scanned relative to the start of the field ionization ramp.  The red and blue curves show the overlap between the field ionization signal at a particular delay and that of the glitch at the beginning of the field ionization ramp for the $s$ and $p$ states respectively.  The time delay of the glitch is converted to the field value at which it arrived using the measured shape of our ramped ionizing field.  This allows us to overlay the glitch scans on the stark map, correlating variations in overlap with particular avoided crossings.
	}
	\label{fig:glitch}
\end{figure*}

Our primary motivation for controlling the shape of the time-resolved field ionization signal is to improve the selectivity of Rydberg state distribution measurements \cite{gregoric_improving_2018}.  In this case our goal is to design a perturbation that best separates the signals that would otherwise overlap with standard SFI.  To do this we alternately excite the two states of interest and measure the time-resolved signal for each state with a given perturbation.  The fitness of a perturbation in separating the two states is then calculated and the perturbations are evolved as before.  

An example optimization is shown in Fig.~\ref{fig:GAsp}.  In this case we sought to separate the signals from the $37s_{1/2}$ and $36p_{3/2}$ states.  Using standard SFI the time-resolved signals from these two states are significantly overlapped as shown in Fig.~\ref{fig:GAsp}(a).  This degree of overlap makes it impossible to quantitatively measure the fraction of population in each state when a mixture is present.  Our GA seeks to move signal from the $s$-state into the region bounded by the two vertical lines.  We chose this particular gate since almost no $p$-state is moved into this region with any of the tested perturbations.  It is therefore likely that the pathway that this part of the $s$-state population takes to ionization is distinct from that of the $p$-state.  This is important since interference at any common avoided crossings encountered by the two states will be affected by the relative phase between the two states in addition to their relative populations.

After running the GA, we are able to move $f_s=27\%$ of the $s$-state into the desired region, while only $f_p = 2\%$  of the $p$-state ends up there.  In an experiment in which a mixture of $s$ and $p$-states is present, we can measure the fraction of signal that ends up in the same gate region, $f_e$.  From these quantities, we can find the fraction of $s$ state in the mixture from 
\begin{equation}
f_{es} = \frac{f_e-f_p}{f_s-f_p}.
\label{eq:fes}
\end{equation}

\section{Probing the Stark Map}

To explore the important features of the Stark map for the $s$ and $p$ states, we took a different approach from that described in Sect.~\ref{sec:insight}.  Here we consider the effect of a minimal perturbation on the shape of the field ionization signals for these two states.  The perturbation in this case is a sudden (5~ns rise time) electric field step of 3~V/cm.  The location of this glitch is varied in 1~ns steps from the beginning to the end of the field ionization ramp, producing a sudden increase in slew rate that is scanned across the ionizing ramp.  At each delay, we record the shape of the field ionization signal and calculate the overlap between this signal and that of the glitch at the beginning of the field ionization ramp.  Slew rate changes at major avoided crossings are likely to shift the path of the electron to ionization which corresponds to a shift in the shape of the ionization signal. The time delay of each glitch is correlated with the field value at which it arrived using the measured time dependent shape of the field ionization ramp.

Figure~\ref{fig:glitch} shows a glitch scan overlaid on the Stark map, with the $s$-state shown in red and the $p$-state in blue.  A decrease in overlap indicates a change in the shape of the field ionization signal.  For the $s$-state we see a large dip in the overlap where the state hits the $n=34$ manifold.  In the $p$-state signal, the first large change in overlap occurs not where the $p$-state hits the $n=33$ manifold, but instead where the $n=33$ and 34 manifolds intersect.  Evidently, the traversal of the avoided crossing between the $p$-state and the $n=33$ manifold remains adiabatic, even with the increased slew rate provided by the glitch.  Clearly, the first avoided crossing plays a very important role in the field ionization process, consistent with early results exploring adiabatic versus diabatic ionization~\cite{jeys_diabatic_1980,noel_classical_2000} as well as early attempts to manipulate the field ionization signal by adjusting the slew rate through the first avoided crossing~\cite{tada_manipulating_2002,gurtler_l-state_2004}.  

Beyond the first large dip, variations in overlap become much more subtle for both states.  This is likely because the population is spread across many states so manipulation of a particular avoided crossing by the glitch only affects a small fraction of the total signal.  It is worth noting that the many dips in overlap occur at different fields for the $s$ and $p$ state scans. This is likely of importance to the ability of the GA to separate the signals from the two states since the perturbation at a particular field will generally have a larger effect on one state than the other.  Finally, we note that the sudden decrease in overlap at high field corresponds to the glitch arriving at the time when the electrons are ionizing.

\section{Dipole-Dipole Interactions}

\begin{figure}
	\centering
	\includegraphics{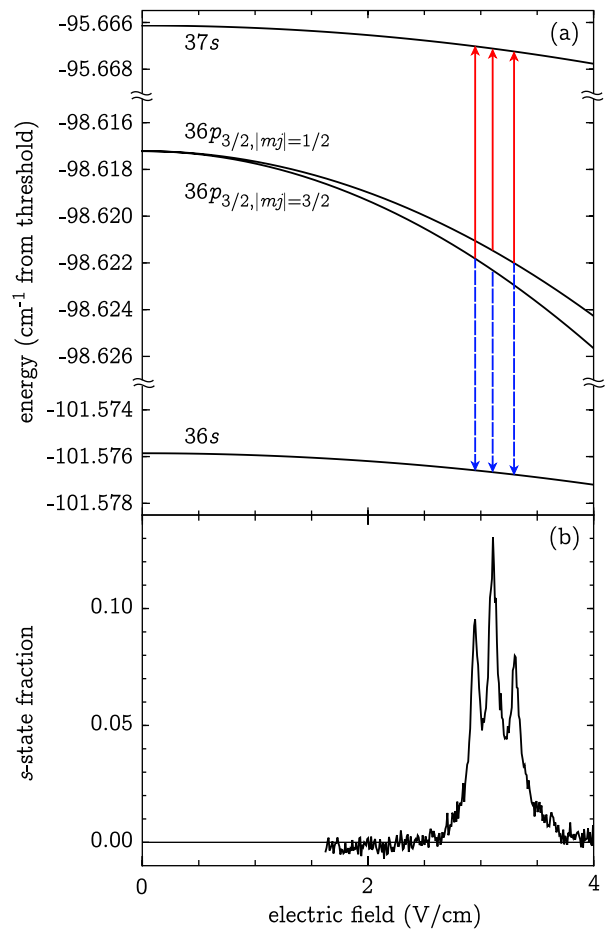}
	\caption{(Color online) 
	Field tuned $36p_{3/2}+36p_{3/2}\rightarrow 36s_{1/2}+37s_{1/2}$ dipole-dipole interaction.  (a) Stark map showing the location of the field tuned resonances.  (b)  Electric field scan showing the fraction of atoms that end up in the $37s$ due to the interaction.
	}
	\label{fig:ppss}
\end{figure}

Separating the $s$ and $p$-state signals allows us to make quantitative measurements of the dipole-dipole energy exchange $36p_{3/2} +36p_{3/2} \rightarrow 36s_{1/2} + 37s_{1/2}$.  This process is tuned into resonance by shifting the states with a small electric field as shown in Fig.~\ref{fig:ppss}(a).  The lowest field resonance is between two atoms in the $|m_j| = 3/2$ state, the highest between two in the $|m_j| = 1/2$ state, and the middle between a pair of atoms, one in each state.


To do this experiment, we first run a GA to find the perturbation that best separates $s$ and $p$ states as described in Sect.~\ref{sec:GAsp}.  We then turn the $s$-state excitation laser off and add a time delay between the Rydberg excitation and DFI to allow for dipole-dipole interaction.  During this window, we use the arbitrary waveform generator to apply a square voltage pulse whose amplitude can be varied.  Rydberg excitation is done at zero field so a mixture of $|m_j|=1/2$ and $3/2$ states are present.  As the amplitude of the voltage pulse is scanned, we see the three dipole-dipole resonances appear.  Figure~\ref{fig:ppss}(b) shows the fraction of population that ends up in the $37s_{1/2}$ state calculated using Eq.~(\ref{eq:fes}) after a 9~$\mu$s interaction.

Observation of this dipole-dipole interaction is possible using standard SFI, but the signal-to-noise ratio is considerably reduced.  In addition, we find that when a mixture of $s$ and $p$ states is present, the SFI signal is not a simple sum of the signals from individually excited states because of interference between common paths taken by the two states during ionization.  This prohibits accurate quantification of the fraction of atoms that end up in the $s$-state due to the dipole-dipole interaction using SFI in this system.  By using DFI to move some of the $s$-state ionization signal to a region that remains inaccessible to the $p$-state, we not only improve the signal-to-noise ratio in our measurement, but are able to accurately quantify the fraction of atoms that have interacted.

\section{Conclusion}

\begin{figure*}
	\centering
	\begin{minipage}{0.9\linewidth}
		\noindent\rule{\textwidth}{0.5pt}
		\begin{algorithmic}[1] 
			\For{$i <$ populationSize} \Comment{Create the initial population of perturbations.}
			\For{$j <$ perturbLength}    
			\State population[i][j] = \Call{RandomFloat}{$-V_{max}$, $V_{max}$} \Comment{Fill each perturbation with random voltages.}
			\EndFor
			\EndFor
			\While{generationCount $<$ generationMax}
			\State \Call{CollectExperimentalData}{population}		\Comment{Collect data with each perturbation.}
			\For{$i <$ populationSize} \Comment{Calculate each perturbation's fitness.}
			\State population[i].fitness = \Call{fitnessFunction}{~}	\Comment{Score each perturbation.}
			\EndFor
			\State population.Sort()		\Comment{Sort perturbations by fitness.}
			\State children[populationSize][perturbLength] \Comment{Create empty population for the next generation.}
			\For{$i < $ eliteCount}
			\State children[i] = population[i] \Comment{Propagate elites directly to the next generation.}
			\EndFor			
			\For{$i=$ eliteCount; $i < $ populationSize}
			\For{$j < 2$}			\Comment{Find two parent perturbations.}
			\State tournament = new List	\Comment{Create a list to store a random sample of our population.}
			\For{$k < $ tournamentSize} \Comment{Select a random sub-population with unique perturbations.}
			\State $r$ = \Call{RandomInt}{0, populationSize}
			\If {$r$ not already used} 
			\State tournament.Add(population[$r$])	
			\EndIf
			\EndFor
			\State tournament.Sort()		\Comment{Find best fitness score in the tournament sub-population.}
			\State parent[$j$] = tournament[0] \Comment{Choose the winner of the tournament as one of our parents.}					
			\EndFor
			\For{$j < $ perturbLength} 	\Comment{Create a new perturbation from our parents using \textit{uniform crossover}.}
			\State $r$ = \Call{RandomFloat}{0, 1}
			\If{$r < 0.5$}			\Comment{Choose randomly with equal weight either parent for each gene.}
			\State child[$j$] = parent[0][$j$]
			\Else
			\State child[$j$] = parent[1][$j$]
			\EndIf
			\EndFor
			\State children[$i$] = child \Comment{Add newly created perturbation to the next generation.}
			\EndFor
			\For{$i < $ populationSize}		\Comment{Randomly select genes for \textit{mutation}.}
			\For{$j < $ perturbLength}
			\State $r$ = \Call{RandomFloat}{0, 1}
			\If{$r < $ mutationRate}			\Comment{Mutate this gene.}
			\State population[$i$][$j$] = \Call{RandomFloat}{$-V_{max}$, $V_{max}$}
			\EndIf
			\EndFor
			\EndFor
			\State population = children		\Comment{Replace the old population with the new population.}
			\State generationCount++
			\EndWhile
		\end{algorithmic}
		\noindent\rule{\textwidth}{0.5pt}
		\caption{Pseudocode for our GA. The relevant section of our C\# source code is available at~\cite{carroll_enhancing_2019}.}		\label{fig:pseudocode}
	\end{minipage}
\end{figure*}

By manipulating the phase evolution of the electron as it traverses hundreds of avoided crossings on the way to ionization, DFI is able to control the shape of the field ionization signal. We demonstrate that DFI can separate field ionization signals that are overlapping when using SFI. This allows us to quantitatively measure a dipole-dipole interaction in which the initial and final states have similar unperturbed pathways to ionization.

Given the complexity of the Stark map in Fig.~\ref{fig:SimulatedGA} and the sensitivity of the avoided crossing traversals to small fields, it is remarkable that there is sufficient coherence for the GA to find good solutions. Previous work has shown that coherence does survive as interference fringes can be measured in the field ionization signal when coherent superposition states are initially excited~\cite{leuchs_quantum_1979,feynman_quantum_2015}. Developing a fitness score for minimizing the interference between two states could be a strategy to improve the ability of the GA to separate the field ionization signals of those states.

Even simple perturbations can alter the shape of the field ionization signal, revealing correlations with the structure of the Stark map as in our glitch experiment. Applications beyond enhancing SFI are possible. For example, quantum control of the electron's pathway to ionization could be useful for applications such as producing an electron beam by ionizing Rydberg atoms~\cite{kime_high-flux_2013}. It is desirable for the electron beam to have a small spread in energy~\cite{mcculloch_field_2017,moufarej_forced_2017}, which could be evolved using a GA.

This work was supported by the National Science Foundation under Grants No. 1607335 and No. 1607377.
\section{Appendix}

In Fig.~\ref{fig:pseudocode} we present a pseudo-code version of the GA. Our GA is implemented in the C\# programming language as a component of our custom data acquisition software on a Microsoft Windows PC. The source code, along with some discussion, is available in the online open source repository GitHub~\cite{carroll_enhancing_2019}. Table~\ref{tab:GAParam} shows typical values of our GA parameters.

Lines 1-5 initialize the population of field ionization perturbations to random voltage values, which are the \textit{genes}. This is the first step shown in Fig.~\ref{fig:GAloopoutline}. The circular loop in Fig.~\ref{fig:GAloopoutline} represents the while loop that spans lines 6-48 in the pseudocode. Once a population is created, we run the experiment and collect our data. Each perturbation is then scored on how well its resulting field ionization signal matches our goal. The perturbations are sorted from best to worst (lines 7-12).

Our GA uses four common strategies: \textit{elitism}, \textit{tournament selection}, \textit{uniform crossover}, and \textit{mutation}~\cite{mitchell_introduction_1998}. These strategies are tuned to strike a balance between converging to a solution and maintaining genetic diversity.

\begin{table}
	\caption{Typical values of GA parameters. The mutation rate is applied to each gene individually. Since we have about 1000 genes, a 2\% chance for each gene is quite significant. We find that our super-elites provide a good balance to this relatively high mutation rate.}\label{tab:GAParam}
	\begin{tabular}{cc}
		\hline 
		\hline 
		parameter & typical value \\ 
		\hline 
		population size & 100 \\ 
		number of elites & 4 \\  
		tournament size & 6 \\ 
		mutation rate & 0.02 \\ 
		\hline
		\hline  
	\end{tabular} 
\end{table}

Since the mating and mutation of population members to produce a new generation involves some amount of randomness, it is possible that the members of the new generation could be worse than the parent generation. To prevent this, we use elitism to propagate the best scoring members of the population directly into the next generation (lines 13-15). 

After the elites have been propagated to the new generation, we fill the rest of the child population by mating pairs of perturbations together. First, tournament selection is used to select a pair of parent perturbations. Two non-overlapping random sets of perturbations are selected and the best perturbation is selected from each (lines 17-27). A child perturbation is created by randomly selecting a parent gene at each locus, a technique referred to as uniform crossover (lines 28-36). We have also tested single point crossover in our GA, in which all genes to the left of a randomly chosen point come from one parent and all genes to the right come from the other. Both techniques produced similar results.

Finally, each gene of each member of the population is subject to a chance of mutation. If a gene is selected for mutation, it is replaced with a random value (lines 38-45). Since even the elites are subject to mutation, this could also result in a new generation that is worse performing than the parent generation. We tested super-elitism, in which we propagated two copies of each elite into the new generation. One copy of each elite was not subject to mutation. We found that this allowed for higher mutation rates and generally faster convergence to a good solution.

We also experimented with dynamic mutation rates, which generally start high when the population is mostly random and no population members are yet a good solution. As the algorithm progresses, the mutation rate is lowered to preserve the presumably improving solutions. We did not find that dynamic mutation significantly improved the performance of our GA.

\bibliography{GAJphysB}

\end{document}